\documentclass[10pt,conference,letterpaper]{IEEEtran}

\usepackage{amsmath,amssymb,bm}
\usepackage{array}
\usepackage{booktabs}
\usepackage{cite}
\usepackage{graphicx}
\usepackage[hidelinks]{hyperref}
\usepackage{microtype}
\usepackage{placeins}
\usepackage[caption=false,font=footnotesize]{subfig}
\usepackage{xcolor}
\usepackage{tikz}
\usepackage{multirow} 
\usetikzlibrary{arrows.meta,calc,fit,patterns,positioning}

\definecolor{EdgeBlue}{HTML}{2F6B9A}
\definecolor{CloudGreen}{HTML}{39815D}
\definecolor{WireOrange}{HTML}{D77A28}
\definecolor{RejectRed}{HTML}{B94A48}
\definecolor{SoftGray}{HTML}{E8EAED}
\definecolor{DarkGray}{HTML}{4A4F55}
\definecolor{CommitGold}{HTML}{D4A72C}

\tikzset{
  paperbox/.style={
    draw=DarkGray!78,
    fill=white,
    rounded corners=1.6pt,
    line width=0.45pt,
    align=center,
    inner sep=2.2pt
  },
  edgebx/.style={paperbox,draw=EdgeBlue!90,fill=EdgeBlue!8},
  cloudbx/.style={paperbox,draw=CloudGreen!90,fill=CloudGreen!8},
  neutralbx/.style={paperbox,fill=SoftGray!55},
  paperarrow/.style={
    -{Latex[length=1.7mm,width=1.25mm]},
    line width=0.65pt,
    draw=DarkGray!88
  }
}

\newcommand{\method}{AsymSpec}
\newcommand{\methodsp}{AsymSpec\ }
\newcommand{\E}{\mathbb{E}}
\newcommand{\Prb}{\mathbb{P}}
\newcommand{\TV}{D_{\mathrm{TV}}}
\newcommand{\pos}[1]{\left[#1\right]_+}

\newcommand{\CorrectionModeFigure}{%
\begin{figure*}[!t]
\centering
\begin{tikzpicture}[
  font=\sffamily\scriptsize,
  panel/.style={draw=DarkGray!38,fill=SoftGray!12,rounded corners=3pt,
                line width=0.45pt},
  title/.style={font=\sffamily\bfseries\scriptsize,
                text=DarkGray,anchor=west},
  smallbox/.style={paperbox,font=\sffamily\scriptsize,
                   minimum height=0.58cm,inner sep=2pt},
  boundbox/.style={smallbox,draw=WireOrange,line width=0.75pt,
                   fill=WireOrange!7},
  samplebox/.style={smallbox,draw=EdgeBlue,line width=0.75pt,
                    dashed,fill=EdgeBlue!6},
  fullbox/.style={smallbox,draw=CloudGreen,line width=0.55pt,
                  double,double distance=0.75pt,fill=CloudGreen!5},
  flow/.style={paperarrow,line width=0.65pt},
  fallback/.style={paperarrow,draw=RejectRed,densely dashed,line width=0.75pt},
  guide/.style={draw=DarkGray!45,densely dotted,line width=0.45pt}
]
\draw[panel] (-8.45,-0.42) rectangle (-2.95,5.55);
\draw[panel] (-2.75,-0.42) rectangle (2.95,5.55);
\draw[panel] (3.15,-0.42) rectangle (8.45,5.55);
\node[title] at (-8.18,5.25) {(a) Residual support and certificate};
\node[title] at (-2.48,5.25) {(b) Three correction modes};
\node[title] at (3.42,5.25) {(c) Online decision};

\node[anchor=east,text=CloudGreen!70!DarkGray,font=\sffamily\bfseries\scriptsize]
  at (-7.72,4.55) {target $p$};
\node[anchor=east,text=EdgeBlue!75!DarkGray,font=\sffamily\bfseries\scriptsize]
  at (-7.72,3.55) {draft $q$};
\node[anchor=east,text=WireOrange!80!DarkGray,font=\sffamily\bfseries\scriptsize]
  at (-7.72,2.55) {Residual $a$};
\draw[DarkGray!55,line width=0.45pt] (-7.58,4.10) -- (-3.26,4.10);
\draw[DarkGray!55,line width=0.45pt] (-7.58,3.10) -- (-3.26,3.10);
\draw[DarkGray!55,line width=0.45pt] (-7.58,2.10) -- (-3.26,2.10);

\foreach \x/\hp/\hq/\ha/\lab in {
 -7.35/0.75/0.50/0.25/$v_1$,
 -6.72/0.64/0.58/0.08/$v_2$,
 -6.09/0.55/0.35/0.20/$v_3$,
 -5.46/0.44/0.48/0.00/$v_4$,
 -4.83/0.34/0.20/0.14/$v_5$,
 -4.20/0.26/0.12/0.14/$v_6$,
 -3.57/0.18/0.06/0.12/tail
}{
  \draw[draw=CloudGreen!85!DarkGray,fill=CloudGreen!17,line width=0.45pt]
    (\x,4.10) rectangle ++(0.18,\hp);
  \draw[draw=EdgeBlue!90!DarkGray,fill=white,line width=0.55pt,
        dash pattern=on 1.6pt off 0.8pt]
    (\x,3.10) rectangle ++(0.18,\hq);
  \ifdim \ha pt>0pt
    \draw[draw=WireOrange!90!DarkGray,pattern=north east lines,
          pattern color=WireOrange!80!DarkGray,line width=0.5pt]
      (\x,2.10) rectangle ++(0.18,\ha);
  \else
    \draw[DarkGray!55,line width=0.65pt] (\x,2.13) -- ++(0.18,0);
  \fi
  \node[anchor=north,text=DarkGray] at ($( \x,2.02)+(0.09,0)$) {\lab};
}
\draw[DarkGray!70,line width=0.55pt]
  (-7.48,1.52) -- (-4.60,1.52);
\draw[DarkGray!70,line width=0.55pt]
  (-7.48,1.52) -- ++(0,0.14)
  (-4.60,1.52) -- ++(0,0.14);
\node[anchor=north,font=\sffamily\bfseries\scriptsize]
  at (-6.04,1.48) {$S=\operatorname{TopK}(p)$};
\draw[guide] (-4.48,1.76) rectangle (-3.28,2.96);
\node[align=center,text=DarkGray] at (-3.88,1.23)
  {omitted tail\\mass $\delta_p$};
\node[smallbox,text width=4.62cm,align=center] at (-5.70,0.62)
  {$a_v=[p(v)-q(v)]_+$,\quad $Z_S=\sum_{v\in S}a_v$\\
   $\TV(r,\hat r)=\dfrac{Z_{\rm tail}}{Z_S+Z_{\rm tail}}
   \leq\dfrac{\delta_p}{Z_S+\delta_p}$.};

\node[boundbox,text width=4.48cm,minimum height=1.06cm,align=left]
  (bounded) at (0.10,4.22)
  {\textbf{Bounded top-$K$}\hfill \textsc{cert. approx.}\\
   Send $(v,p(v))_{v\in S}$ + tail mass;\quad bytes $O(K)$\\
   Query $q(v)$; reserve $\epsilon_j\leq\epsilon_{\rm rem}$.};
\node[samplebox,text width=4.48cm,minimum height=1.06cm,align=left]
  (sampled) at (0.10,2.75)
  {\textbf{Exact proposals}\hfill \textsc{exact; dashed}\\
   Send ordered proposals $(Y,p(Y))$;\quad entries $O(N)$\\
   Accepted proposal follows $r$; cap trials, then fallback.};
\node[fullbox,text width=4.48cm,minimum height=1.06cm,align=left]
  (full) at (0.10,1.28)
  {\textbf{Full distribution}\hfill \textsc{exact; double}\\
   Send full $p$;\quad bytes $O(|V|)$\\
   Reconstruct residual using queryable $\mathsf{Q}_i$.};
\draw[fallback] (bounded.south) -- (sampled.north);
\draw[fallback] (sampled.south) -- (full.north);

\node[neutralbx,text width=4.05cm,minimum height=0.96cm,align=center]
  (inputs) at (5.80,4.55)
  {\textbf{Receive nested target top-$K$}\\
   query retained $q(v)$ on the returned support};
\node[boundbox,minimum width=2.00cm,minimum height=0.78cm,align=center]
  (quality) at (4.52,3.18)
  {\textbf{Compute}\\residual-TV\\certificate};
\node[samplebox,minimum width=2.00cm,minimum height=0.88cm,align=center]
  (deadline) at (7.08,3.18)
  {\textbf{Compare}\\with remaining\\request budget};
\node[smallbox,text width=4.05cm,minimum height=1.12cm,align=center]
  (choice) at (5.80,1.90)
  {\textbf{Certificate-driven progression}\\
   fits $\Rightarrow$ reserve + bounded draw\\
   fails $\Rightarrow$ expand $K$ or recover exactly};
\node[edgebx,minimum width=1.20cm,minimum height=0.48cm] (outk) at (4.20,1.02)
  {commit};
\node[paperbox,draw=EdgeBlue,dashed,minimum width=1.35cm,minimum height=0.48cm]
  (outs) at (5.82,1.02) {expand $K$};
\node[paperbox,draw=CloudGreen,double,double distance=0.65pt,
      minimum width=1.15cm,minimum height=0.48cm]
  (outf) at (7.38,1.02) {exact};
\draw[flow] (inputs) -- (quality);
\draw[flow] (inputs) -- (deadline);
\draw[flow] (quality) -- (choice);
\draw[flow] (deadline) -- (choice);
\draw[flow] (choice) -- (outk);
\draw[flow] (choice) -- (outs);
\draw[flow] (choice) -- (outf);
\node[anchor=south,align=center,text width=4.85cm,text=RejectRed!85!DarkGray,
      font=\sffamily\scriptsize]
  at (5.80,-0.34)
  {$\sum_j\epsilon_j\leq\epsilon_{\rm req}$ for bounded draws;\\
   exact recovery contributes zero TV charge.};
\end{tikzpicture}
\caption{Residual-certified correction. Panel (a) separates the
covered residual support from omitted tail mass; the bars are schematic rather
than measurements. Panel (b) distinguishes bounded top-$K$ correction
from the two exact paths using both text and border style. Panel (c) makes the
realized certificate, rather than a heuristic target-coverage threshold, the
decision boundary for expansion or exact recovery.}
\label{fig:correction-mode-selector}
\end{figure*}
}

\newif\ifshowauthors
\showauthorstrue

\title{AsymSpec: Efficient Cloud--Edge Speculative Decoding over Asymmetric Networks}
\ifshowauthors
  \IEEEoverridecommandlockouts
\author{%
  \IEEEauthorblockN{Guotao Yang, Hao Chen, Rui Guo, Xinyu Li, Liang Zheng,
  Sheng Chen, Yitao Hu\textsuperscript{*}, and Keqiu Li}
  \IEEEauthorblockA{Tianjin University, China}
  \thanks{\textsuperscript{*}Corresponding author: Yitao Hu
  (e-mail: \texttt{yitao@tju.edu.cn}).}
}
\else
\author{}
\fi

\begin{document}
\maketitle

\begin{abstract}
Cloud--edge speculative decoding places a lightweight draft model at an edge gateway and a higher-quality target model in the cloud, but inserts communication into every speculative block. Under a constrained uplink, candidate messages may queue while the verifier is idle. Stop-and-wait scheduling leaves edge compute underutilized; optimistic same-request runahead can waste work when a rejection or an unexpected bonus token invalidates dependent drafts. We present \method, which addresses uplink-gated verification and invalid dependent work with two corresponding mechanisms. Its asymmetric verification protocol keeps the common-path acceptance upload compact and moves richer, rejection-only correction information to the downlink. A total-variation (TV) certificate for the residual distribution determines whether a small target top-$K$ response suffices; if not, the protocol progressively escalates through proposal-based exact recovery before falling back to the full distribution. Its confirmed-prefix pipeline exposes only independent, valid requests to the edge scheduler and lets the cloud re-batch arrived blocks, hiding verification waits when another confirmed-prefix request is ready without using same-request runahead. Across three draft--target pairs, two workloads, and three asymmetric network profiles, our end-to-end evaluation shows that \method{} delivers 2.82--28.03$\times$ the output-token throughput of the strongest baseline.
\end{abstract}


\section{Introduction}

Large language models (LLMs) now power
dialogue~\cite{thoppilan2022lamda}, code generation~\cite{chen2021evaluating},
knowledge-intensive question answering~\cite{brown2020language}, and complex
reasoning~\cite{wei2022chain}. A growing class of applications wraps these
models in agentic loops that plan, invoke external tools, and incorporate the
returned observations before generating again~\cite{yao2023react}. Unlike a conventional query that produces one response and exits, an agent task repeatedly alternates model generation, tool execution, and feedback incorporation, and each round initiates another autoregressive generation phase. Because
every token can be sampled only after its predecessor has been fixed, the
serial cost that is modest for a single reply compounds across the many rounds
of an agent interaction. Multi-round LLM workloads therefore amplify the
throughput cost of autoregressive decoding.

One way to relieve this bottleneck is speculative
decoding~\cite{leviathan2023,chen2023}. A lightweight draft model proposes a
short continuation, and a high-quality target model verifies multiple candidate
positions in parallel. The longest accepted prefix is committed; any rejection
is resolved with a target-guided correction before the next round. In effect,
\emph{several serial target decoding steps are replaced by one parallel
verification pass, reducing the number of target invocations per output
token.} Crucially, the standard acceptance test and residual correction rule
preserve the target model's output distribution exactly, so the acceleration
comes without sacrificing output quality~\cite{leviathan2023,chen2023}.

The gains of speculative decoding within a single request motivate a natural
follow-up question: when many requests are served concurrently, can the
draft--verify split also be distributed across machines? Recent systems place
draft generation on an edge device and target verification in the
cloud~\cite{dssd,specedge,pipesd,zhengcomm,zhengfast}. This placement
matches a two-sided resource asymmetry: edge devices usually cannot host a
high-quality target model but can run a lightweight draft model, while cloud
servers provide the GPU memory and compute that verification demands. The edge
handles inexpensive proposal generation, the cloud concentrates on verification
across concurrent requests, and otherwise idle edge capacity is put to use.
\emph{Cloud--edge execution is therefore attractive, but nontrivial:} its
benefit depends on how effectively the system converts asymmetric network and
compute resources into output tokens~\cite{distance,llmcad}. The cloud
target must still verify every accepted candidate, and rejection correction is
either exact or explicitly bounded.

\begin{figure}[!t]
\centering
\includegraphics[pagebox=cropbox,width=0.48\textwidth]{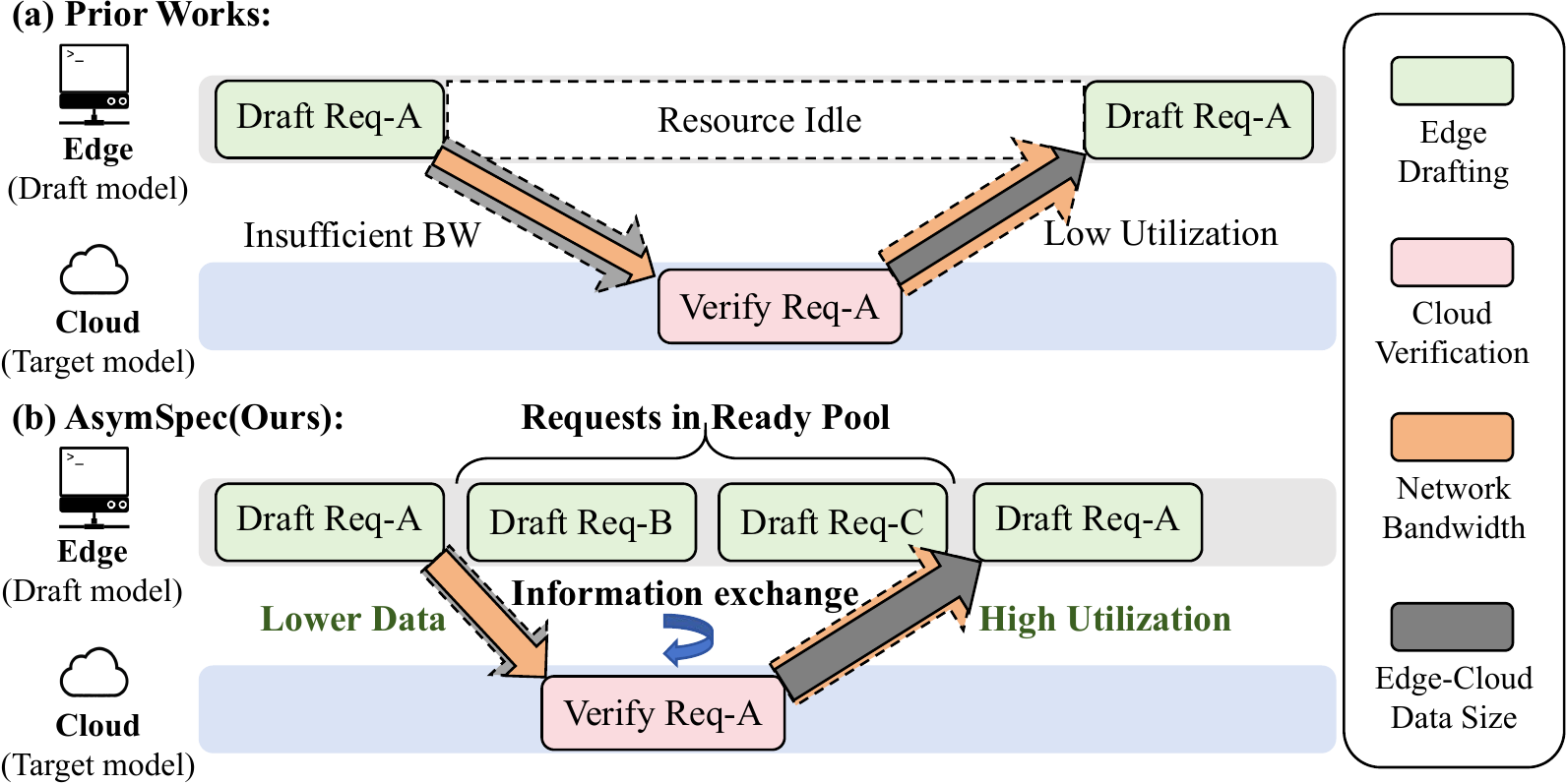}
\caption{Conceptual comparison of prior cloud--edge speculative serving and
\method. Prior designs can leave edge resources idle while a request is
verified. \method{} reduces common-path communication and drafts independent
requests from the ready pool during the verification wait, improving edge
utilization without drafting descendants of an unresolved block.}
\label{fig:intro-overview}
\end{figure}

Two concrete obstacles stand in the way of this goal.

\textbf{Limitation 1: the verification path does not match asymmetric
bandwidth.}
Every speculative block must traverse the constrained uplink before the cloud
can verify it, although common-path acceptance needs only candidate tokens and
their draft probabilities. Rejection is less frequent but may require much
richer target-side information for correction, which can instead use the
stronger downlink. Treating both outcomes as one symmetric exchange therefore
either inflates every common-path upload or returns excessive correction state.
This mismatch can queue block uploads and starve the verifier while downlink
capacity remains available. Moreover, a compact correction response still
needs a residual-specific guarantee: target top-$K$ coverage alone is
insufficient because the draft may match the target on the head while their
positive difference lies in the tail. Section~\ref{sec:motivation-uplink}
characterizes this directional load condition.

\textbf{Limitation 2: dependent runahead can turn overlap into invalid work.}
The next draft round of a request must begin from the tokens confirmed by its
current verification. Waiting for that result creates a draft-compute bubble.
Optimistic runahead hides the wait by drafting from an unconfirmed prefix, but
a rejection then invalidates the dependent tokens, their transmissions, and
possibly their cloud verification~\cite{pipeinfer,amusd,pearl,pipesd}.
Prior systems already interleave requests and batch
verification~\cite{specedge,cosine,wisp}. The remaining dependency question is
whether overlap can be restricted
to the \emph{confirmed frontier}: while one request awaits verification,
another independent request may be ready without creating any descendant of an
unresolved block.

To address these two gaps, we present \textbf{\method{}}, which pairs a communication
mechanism with a scheduling invariant. On the communication side, it
progressively expands target support under a certificate for the realized
residual distribution, composes each bounded correction into a request-level
fidelity budget, and falls back to exact recovery when the certificate is
insufficient. On the scheduling side, it admits only confirmed-prefix requests
and re-batches their arrived blocks at the cloud. We call a speculative block
\emph{open} from the start of edge drafting until it is committed or aborted;
\method{} allows at most one such block per request. This invariant forbids dependent runahead rather than merely rescheduling it; when another request is ready, the scheduler uses that confirmed-prefix work to cover the verification wait.
Figure~\ref{fig:intro-overview}
contrasts this confirmed-prefix pipeline with prior request-internal execution:
the edge uses independent ready requests to cover verification waits while the
common path carries less data.

This paper makes three contributions:
\begin{itemize}
  \setlength{\itemsep}{0pt}
  \item \textbf{Problem analysis.} We quantify the directional payload imbalance on the verification path and characterize the validity boundary between idle waiting and request-internal work that a rejection or unseen bonus token in an ancestor block can invalidate.
  \item \textbf{Protocol and system design.} We specify \method, combining an
  acceptance-sufficient common-path uplink with rejection-only progressive
  correction. The
  protocol certifies the residual-distribution error rather than relying on
  target top-$K$ coverage alone, and composes each bounded correction into a
  request-level fidelity budget. A confirmed-prefix cross-request pipeline then
  decouples edge drafting from cloud verification batches without generating
  descendants of unresolved blocks.
  \item \textbf{Experimental evaluation.} We evaluate \method{} across three
  draft--target model pairs, two workloads, and three asymmetric network
  profiles. Relative to the strongest baseline, \method{} delivers
  \textbf{2.82--28.03$\times$} the output-token throughput.
\end{itemize}

\section{Background}

\subsection{LLM Inference and the Decoding Bottleneck}

An autoregressive large language model (LLM) generates an output by repeatedly
predicting the next token conditioned on the confirmed
prefix~\cite{brown2020language}. Its inference has a parallel \emph{prefill} phase
over the input prompt and a sequential \emph{decode} phase over the output.
During decoding, a key--value (KV) cache reuses states from the confirmed
prefix instead of recomputing them~\cite{vllm,distserve}. This cache reduces
redundant computation within each step, but it does not remove the dependency
across steps: token $x_{n+1}$ must be selected before the model can generate
$x_{n+2}$. Conventional decoding therefore invokes the full model once for
each output token, leaving little opportunity to exploit parallelism across
positions within one sequence.

% Speculative decoding targets this serial decode bottleneck without changing the
% target model's autoregressive semantics. A lightweight model first supplies
% provisional future tokens, allowing one target-model forward pass to evaluate
% several candidate positions in parallel. Because the draft and target
% distributions differ, however, parallel scoring alone cannot determine which
% tokens may be committed. Speculative sampling closes this gap with a
% target-guided acceptance-and-correction rule that converts the parallel scores
% into the same output distribution as ordinary target
% decoding~\cite{leviathan2023,chen2023}. The next subsection formalizes this
% proposal--verification connection.

\subsection{Speculative Decoding}

Speculative decoding targets this serial decode bottleneck without changing the
target model's autoregressive semantics. A lightweight draft model first
supplies provisional future tokens, allowing one target-model forward pass to
evaluate several candidate positions in parallel. Because the draft and target
distributions differ, however, parallel scoring alone cannot determine which
tokens may be committed. Speculative sampling closes this gap with a
target-guided acceptance-and-correction rule that converts the parallel scores
into the same output distribution as ordinary target
decoding~\cite{leviathan2023,chen2023}.

Let $x_{1:n}$ denote the confirmed prefix. The draft model extends it with a
block of provisional tokens. At speculative position $i$, let $q_i$ denote the
actual post-processed draft distribution conditioned on the confirmed prefix
and preceding draft candidates, and let $p_i$ denote the target distribution
evaluated at the same candidate position. Both are normalized categorical
distributions over the same vocabulary $\mathcal{V}$, represented by the canonical
finite-precision values actually used for sampling. The endpoints bind each
block to tokenizer, model, and
sampling-configuration versions; each endpoint must reproduce its own
distribution exactly under that configuration. We focus on sampling; greedy
decoding can be handled as a special case. The draft samples
$\tilde{x}_i\sim q_i$. The target accepts it with probability
\begin{equation}
a_i=\min\!\left(1,
\frac{p_i(\tilde{x}_i)}{q_i(\tilde{x}_i)}\right).
\label{eq:accept}
\end{equation}

At the first rejected position, standard speculative sampling draws the
correction token from the residual distribution
\begin{equation}
r_i(v)=\frac{\pos{p_i(v)-q_i(v)}}{Z_i},\quad
Z_i=\sum_u\pos{p_i(u)-q_i(u)}.
\label{eq:residual}
\end{equation}
For normalized $p_i$ and $q_i$, $Z_i=\TV(p_i,q_i)$, where $\TV$ denotes
total-variation distance, and $Z_i$ equals the rejection probability for a
candidate sampled from $q_i$. If $Z_i=0$, the correction path is never entered
at that position. Replacing~\eqref{eq:residual} with an uncertified compressed
approximation can therefore change the target distribution.

Ignoring end-of-sequence tokens and maximum-length truncation, a block of draft
depth $\gamma$ commits every consecutively accepted candidate and one correction
or bonus token. On the all-accepted path, the bonus is sampled with fresh randomness
from the full target distribution after the accepted block. Let $C_\gamma$ denote
the number of tokens committed by a block of draft depth $\gamma$. If $\alpha_i$
is the conditional acceptance probability at position $i$, then
\begin{equation}
\E[C_\gamma]=1+\sum_{j=1}^{\gamma}
\prod_{k=1}^{j}\alpha_k.
\label{eq:yield}
\end{equation}
Under an approximation in which position-wise acceptance events are independent
and identically distributed with common probability $\alpha$, $\alpha^\gamma$
estimates only the probability that all draft candidates survive; it is not the
total block yield.

\subsection{Cloud--Edge Collaborative Speculative Inference}

Cloud--edge speculative inference places the lightweight draft model at the edge and the larger target model in the cloud~\cite{dssd,specedge,pipesd,zhengcomm,zhengfast}. This placement
matches their resource requirements: the edge handles inexpensive proposal
generation, whereas the cloud provides the memory and compute required for
target verification. In each round, the edge drafts from a confirmed prefix
and uploads a candidate block; the cloud verifies the block and returns the
committed result. Both endpoints then advance to the same prefix.

The split places the draft distribution $q_i$ at the edge and the target
distribution $p_i$ in the cloud. Given a shared confirmed prefix and sampling
configuration, the cloud can evaluate acceptance using only
\begin{equation}
m_i^{up}=(\tilde{x}_i,q_i(\tilde{x}_i)),
\label{eq:minmsg}
\end{equation}
plus compact metadata. Exact correction, by contrast, may depend on
$q_i(v)$ for tokens other than $\tilde{x}_i$. The common acceptance path and
the conditional correction path can therefore impose substantially different
communication requirements.

Round-wise synchronization creates a separate dependency between drafting and
verification. Waiting for the cloud response leaves the edge idle, whereas
same-request runahead from an unresolved prefix can be invalidated by a
rejection or an unseen bonus token~\cite{pipeinfer,amusd,pearl}. An independent
request with a confirmed prefix does not have this dependency. These
path-dependent communication and work-validity considerations motivate the
next section.

\section{Motivation}

Cloud--edge speculative decoding couples model execution with network
communication and round-wise synchronization, so its bottlenecks cannot be
inferred from model compute alone. We first quantify the end-to-end latency
composition and then analyze the two dominant sources of inefficiency revealed
by this breakdown.

\subsection{End-to-End Bottleneck Breakdown}
\label{sec:motivation-breakdown}

\begin{figure}[!t]
  \centering
  \includegraphics[width=0.923\columnwidth]{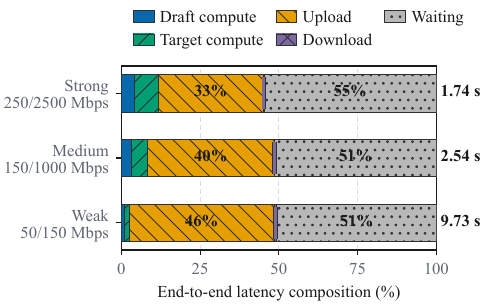}
  \caption{End-to-end latency breakdown under progressively weaker asymmetric
  network conditions. Communication (upload and download) and verification
  waiting together dominate total latency, while local draft and target compute
  remain a minor fraction. Segment labels indicate per-component shares; values
  at the bar ends report total latency.}
  \label{fig:network-breakdown}
\end{figure}

We instrument a representative setup (Qwen3 4B/32B on GSM8K) and measure its
latency breakdown under progressively weaker asymmetric network conditions.
Figure~\ref{fig:network-breakdown} shows that total latency rises from
1.74\,s to 2.54\,s and 9.73\,s as the network profile weakens. The upload
share grows from 33.3\% to 40.2\% and 46.0\%, while the waiting component
remains the largest at 50.6--54.7\%. Together, communication and waiting
account for over 85\% of end-to-end latency under all conditions, far
exceeding local draft and target compute.

The breakdown exposes two distinct bottlenecks. First, every speculative block
must traverse the uplink before verification, making the upload a substantial
part of the critical path. Second, the large waiting component indicates that
round-wise synchronization leaves considerable edge compute unproductive. The
following two subsections analyze these communication and scheduling
bottlenecks, respectively.

\subsection{Limitation 1: The Verification Path Does Not Match
Asymmetric Bandwidth}
\label{sec:motivation-uplink}

Existing cloud--edge speculative verification can place a disproportionate
amount of traffic on the constrained uplink. Every speculative block must
arrive at the cloud before verification can begin, and an exchange that uploads
a vocabulary-sized distribution for each candidate position incurs a payload
that grows with both the candidate count and the model vocabulary. The reverse
message carrying the verification result is substantially smaller.

\begin{table}[t]
\centering
\caption{Directional payloads of a full-distribution verification exchange.
$N$ is the total number of candidate positions in a transmitted batch.
The corresponding download payloads for $N=4$, $32$, and $256$ are
64\,B, 512\,B, and 4,096\,B, respectively.}
\label{tab:directional-payload}
\setlength{\tabcolsep}{3.2pt}
\begin{tabular}{lccccc}
\toprule
\multirow{2}{*}{Model} & \multirow{2}{*}{$|\mathcal{V}|$}
& \multicolumn{3}{c}{Upload payload (MiB)}
& \multirow{2}{*}{Up/down} \\
\cmidrule(lr){3-5}
& & $N=4$ & $N=32$ & $N=256$ & \\
\midrule
Qwen3     & 151,936 & 1.159 & 9.273 & 74.188 & 18,992$\times$ \\
GLM-4     & 151,552 & 1.156 & 9.250 & 74.000 & 18,944$\times$ \\
Llama 3.1 & 128,256 & 0.979 & 7.828 & 62.625 & 16,032$\times$ \\
\bottomrule
\end{tabular}
\end{table}

As shown in Table~\ref{tab:directional-payload}, even a batch containing only
four candidate positions produces a 0.98--1.16\,MiB upload, compared with a
64\,B download. At $N=256$, the upload reaches 62.63--74.19\,MiB, while the
download remains 4,096\,B. The upload-to-download ratio is therefore
16,032--18,992$\times$ across the three models. Increasing $N$ from 4 to 256
increases both the candidate count and the upload payload by 64$\times$,
showing that the uplink cost scales linearly with $N|\mathcal{V}|$.

This vocabulary-scale upload is disproportionate to the information needed on
the common verification path. As summarized in~\eqref{eq:minmsg}, acceptance
testing at each candidate position requires only the sampled token, its draft
probability, and compact metadata, whereas richer distribution information is
needed only when rejection requires correction. A full-distribution exchange
therefore places this conditional information cost on every recurring uplink
message, which can gate verification even when cloud compute and downlink
capacity remain available.

\subsection{Limitation 2: Dependent Runahead Creates Invalid Work}
\label{sec:motivation-validity}

\begin{figure}[!t]
  \centering
  \includegraphics[width=\columnwidth]{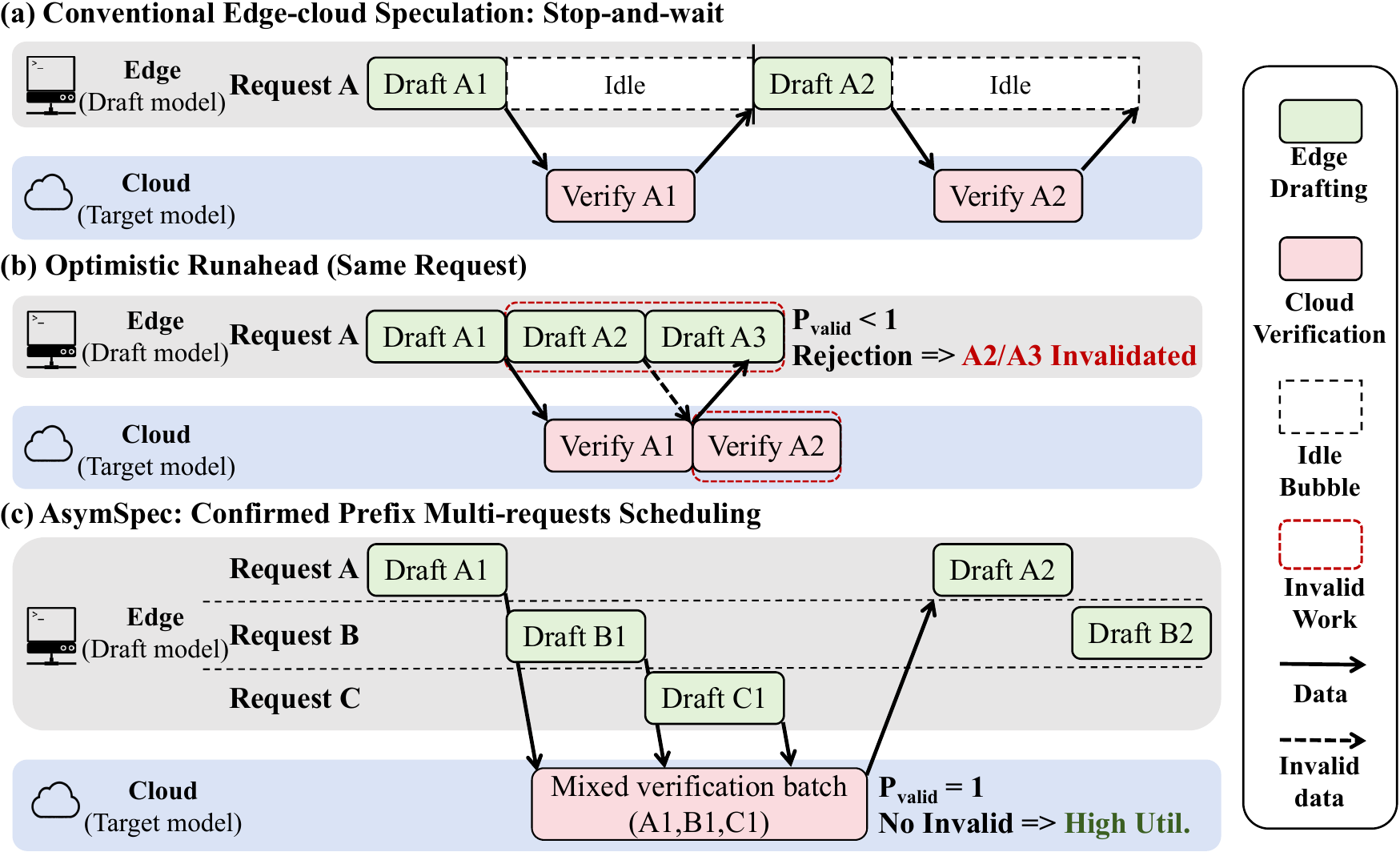}
  \caption{Execution alternatives during verification. (a) Stop-and-wait
  leaves the edge idle until A1 is confirmed. (b) Same-request runahead drafts
  A2 and A3 from an unresolved prefix, so their work may be invalidated when A1
  changes the committed prefix. (c) Blocks drafted from independent confirmed
  prefixes provide a dependency-free comparison. The timeline is schematic and
  not to scale.}
  \label{fig:timeline}
\end{figure}

Existing request-internal execution faces two unsatisfactory choices while a speculative block is being verified. Stop-and-wait preserves the confirmed prefix but leaves the edge idle, whereas same-request runahead fills the wait with descendants of an unresolved block whose results may later become invalid. Figure~\ref{fig:timeline}(a) and (b) illustrate these two cases, respectively.

Under stop-and-wait in Figure~\ref{fig:timeline}(a), the edge drafts A1 and
then waits for upload, cloud queueing and verification, downlink, and commit
before beginning A2. Let $T_{edge}$ denote the drafting time and $T_{off}$ the
interval during which the request is off the edge critical path. The edge
utilization attributable to this request is bounded by
\begin{equation}
U_{edge}^{single}\leq
\frac{T_{edge}}{T_{edge}+T_{off}}.
\label{eq:singleutil}
\end{equation}
As communication and verification delays increase, $T_{off}$ dominates the
denominator and the idle interval in Figure~\ref{fig:timeline}(a) expands.

Same-request runahead, shown in Figure~\ref{fig:timeline}(b), increases raw
utilization by drafting A2 and A3 before A1 is resolved. However, these
descendants are conditioned on the assumption that A1 commits exactly the
prefix used to draft them. Let $P_{\mathrm{valid}}$ denote the probability
that the dependent drafts A2--A3 in Figure~\ref{fig:timeline}(b) remain valid.
If A1 contains $\gamma$ candidates and each candidate is accepted with
probability $\alpha$, approximating the acceptance events as independent gives
\begin{equation}
P_{\mathrm{valid}}\leq\alpha^\gamma.
\label{eq:runahead}
\end{equation}
For $0<\alpha<1$, the upper bound decreases exponentially with $\gamma$, but it may still be loose because, even when every candidate in A1 is accepted, the target may append an unseen bonus token that changes the prefix from which A2 should have been drafted. A rejection or unseen bonus can therefore invalidate dependent draft tokens,
their transmissions, associated verification work, and KV suffixes.

Together, Figure~\ref{fig:timeline}(a) and (b) expose the limitation of request-internal execution. Stop-and-wait preserves validity but leaves the edge idle, whereas runahead increases utilization by producing work whose validity depends on an unresolved block. Raw utilization can consequently increase without a corresponding improvement in throughput.

\subsection{Takeaways}
The analysis yields two conclusions: (1) the verification exchange should
match asymmetric bandwidth and event frequency, avoiding vocabulary-scale
traffic on the recurring uplink path when richer information is needed only
conditionally; and (2) scheduling efficiency should be measured by valid
committed work rather than raw utilization, since stop-and-wait leaves compute
idle while dependent runahead may spend resources on work that is later
invalidated.

\section{AsymSpec Design}
\label{sec:design}

\subsection{Design Overview}

\methodsp consists of an edge gateway and a cloud verifier. The edge maintains confirmed request prefixes, permits at most one open block per request, generates draft blocks, and retains the associated draft state. The cloud collects asynchronously arrived blocks, dynamically forms verification batches across requests, and verifies them with the target model. The design combines an asymmetric verification protocol with a request-decoupled execution pipeline.

During execution, the edge selects a ready request, generates a draft block, and uploads token--scalar pairs $(\tilde{x}_i,q_i(\tilde{x}_i))$ with compact metadata. The cloud dynamically batches incoming blocks and verifies their candidate positions in parallel. On the all-accepted path, it returns the accepted length and a bonus token; at the first rejection, it invokes progressive correction using top-$K$ support, exact proposals, or full fallback. After the result is resolved, the edge and cloud commit the same suffix, and the request becomes ready for its next block.

\subsection{Asymmetric Verification Protocol}

\emph{Insight and approach.}
Acceptance and correction differ in when they are required and in their
information demands. Every speculative block must be checked, but acceptance
requires only each candidate token and its draft probability; a rejection is
conditional and may require richer target-side information for correction.
\method{} therefore
uploads only acceptance-sufficient token--probability pairs and returns
correction information on demand over the downlink. After a rejection, it
starts with a small target top-$K$. Using the retained draft distribution, the edge bounds the TV distance between the support-truncated correction distribution defined by these $K$ tokens and the standard exact residual distribution. We
call this bound the \emph{residual-distribution TV certificate}. The protocol
expands $K$ until the bound fits the request's error budget and otherwise falls
back to an exact path.

\subsubsection{Acceptance-Sufficient Uplink}

Equation~\eqref{eq:accept} shows that the cloud requires only
$(\tilde{x}_i,q_i(\tilde{x}_i))$ for candidate $i$. A depth-$\gamma$ block therefore
uploads $O(\gamma)$ token--scalar pairs instead of
$O(\gamma|\mathcal{V}|)$ draft probabilities. The full draft distribution
remains at the edge through a queryable draft state $\mathsf{Q}_i$, defined by
$\mathsf{Q}_i(v)\triangleq q_i(v)$ for every $v\in\mathcal{V}$; after a
rejection, the edge evaluates $\mathsf{Q}_i(v)$ only for tokens selected by the
cloud's correction response.
This converts an unconditional distribution upload into a demand-driven
downlink support query.

The target evaluates all candidate positions in parallel and applies
\eqref{eq:accept} from left to right. An all-accepted block follows the compact
result path, whereas the first-rejection path retains the target distribution
and accepted-prefix state until one correction token is chosen. If $\mathsf{Q}_i$
cannot reproduce the agreed draft probabilities, the block aborts. Every
transmitted acceptance scalar, certificate statistic, and exact-recovery
probability uses a lossless encoding of the canonical normalized
finite-precision value used for sampling; lossy encodings are out of scope.
Block identifiers bind every response to the request, base-prefix version,
candidate position, and sampling configuration that produced it. This binding
makes delayed or duplicated messages detectable and prevents a correction from
being applied after the request has advanced. Retransmission is idempotent: the
endpoints reuse cached acceptance decisions, target supports, and random
outcomes rather than sampling again. These rules do not change the underlying
acceptance law, but they make the distributional argument meaningful in an
asynchronous implementation where messages can be retried or arrive late.

\subsubsection{Certified Progressive Correction}
\label{sec:bounded-topk}

At the rejected position, write $p$, $q$, $r$, and $Z$ for the corresponding
$p_i$, $q_i$, $r_i$, and $Z_i$. The cloud first returns a small target top-$K$.
The edge combines this support with local draft probabilities to construct a
bounded residual approximation. Crucially, target top-$K$ coverage alone is
not a distribution-fidelity guarantee: $p$ and $q$ may agree on most of the returned head
while the positive residual $[p-q]_+$ lies in the omitted tail. \method{}
therefore certifies error in the \emph{correction distribution} itself.

Order tokens by decreasing $p(v)$, break ties by token ID, and let
$S=\mathrm{TopK}(p)$. From the returned $(v,p(v))$ pairs and local $q(v)$
lookups, the edge computes
\begin{equation}
\begin{aligned}
\delta_p&=1-\sum_{v\in S}p(v),
&a_v&=\pos{p(v)-q(v)},\\
Z_S&=\sum_{v\in S}a_v .
\end{aligned}
\label{eq:topkstats}
\end{equation}
For $Z_S>0$, define $\hat r(v)=a_v/Z_S$ on $S$ and zero outside $S$.

\textbf{Proposition 1 (Residual-TV certificate).}
Let $Z_{tail}=\sum_{v\notin S}\pos{p(v)-q(v)}$. Then
\begin{equation}
\TV(r,\hat r)=\frac{Z_{tail}}{Z_S+Z_{tail}}
\leq\underbrace{\frac{\delta_p}{Z_S+\delta_p}}_{\bar\epsilon_K}.
\label{eq:topkbound}
\end{equation}
Restricting $r$ to $S$ moves exactly
$Z_{tail}/(Z_S+Z_{tail})$ probability mass, and
$Z_{tail}\leq\delta_p$, which proves the bound. The certificate is computable
at the edge from the returned target support and retained draft state, so the
downlink support can be expanded until the computable certificate meets the
requested error budget. A valid
implementation must accumulate these statistics exactly or conservatively
round $\delta_p$ upward and $Z_S$ downward, so numerical error cannot
understate the certificate.

For correction $j$ following pre-correction history $h$, set
$\epsilon_j(h)=\bar\epsilon_{K_j(h)}$ when bounded correction uses selected
support size $K_j(h)$, and set $\epsilon_j(h)=0$ when exact correction is used.
A bounded draw is allowed only when $Z_S>0$ and $\epsilon_j(h)$ fits the
request's remaining budget.
Otherwise, the cloud extends the same ordered support; increasing $K$ cannot
increase $\delta_p$ or decrease $Z_S$ and therefore cannot worsen the
certificate. A failed certificate ultimately enters an exact mode rather than
relaxing fidelity.

\textbf{Corollary 1 (Request-level budget).}
For a finite request under a common fixed stopping and abort policy and with a
request-level TV budget $\epsilon_{req}$, let $\mathcal{S}_{exact}$ and
$\mathcal{S}_{bounded}$ denote the distributions of completed model-token
traces under exact correction throughout and under the budgeted protocol,
respectively. Suppose that, for every correction $j$ and every reachable
pre-correction history $h$, the TV distance between the exact and implemented
conditional correction distributions is at most $\epsilon_j(h)$. For an online
path, write $h_{<j}$ for its history before correction $j$. If every online path
satisfies
\begin{equation}
\sum_j\epsilon_j(h_{<j})\leq\epsilon_{req},
\label{eq:reqbudget}
\end{equation}
then
$\TV(\mathcal{S}_{exact},\mathcal{S}_{bounded})\leq\epsilon_{req}$.
A coupling that charges the first differing correction proves this implication.
Each realized bounded correction is charged once; exact corrections consume
zero budget. If abort is externally observable, it is included as an output
symbol in both distributions.

\subsubsection{Exact Recovery}

When the certificate is insufficient, immediately sending the full target
distribution would forfeit the communication benefit. \method{} first uses a
proposal-based exact path: with randomness independent of the triggering
rejection and previous trials, the cloud draws $Y\sim p$ and sends only
$(Y,p(Y))$, while the edge retrieves $q(Y)$ and accepts with
\begin{equation}
\begin{aligned}
b(v)&=
\begin{cases}
\pos{1-\dfrac{q(v)}{p(v)}}, & p(v)>0,\\
0, & p(v)=0,
\end{cases}\\
\Prb(Y=v,\mathrm{accept})
&=p(v)b(v)=\pos{p(v)-q(v)} .
\end{aligned}
\label{eq:sampleaccept}
\end{equation}
The uniform variate used for acceptance is independent of $Y$ and all other
protocol randomness and is compared against $b(Y)$. Conditioned on success,
the proposal is therefore distributed
exactly as the residual $r$ in~\eqref{eq:residual}; this instantiates the
standard rejection-sampling identity as a two-endpoint correction protocol. If the protocol attempts $N_{\mathrm{prop}}$ mutually independent proposals,
\[
\Prb(\text{all $N_{\mathrm{prop}}$ proposals fail}\mid p,q)
=(1-Z)^{N_{\mathrm{prop}}}.
\]
The first accepted proposal is exact; if all attempts fail, an independent full
residual draw is also exact. Repeated delivery of the same proposal attempt
reuses its cached outcome rather than resampling. This cap is important
because high draft--target
agreement makes rejection rare but can make proposal acceptance, whose mass is
$Z$, small once rejection occurs.

Bounded top-$K$, proposal-based exact recovery, and full fallback transmit $O(K)$, $O(N_{\mathrm{prop}})$, and $O(|\mathcal{V}|)$ entries across one correction, respectively.
Thus the protocol pays for a full distribution only when neither a certified
support nor a bounded number of exact proposals suffices.

The three modes form a staged runtime decision rather than a fixed global
choice of $K$: start from a small support, expand it monotonically when the
certificate is too loose, and reserve exact recovery for the remaining cases.
Because residual-distribution error rather than target coverage alone determines sufficiency, the same $K$ may pass one rejection and fail another. Per-event certification thus avoids a workload-specific threshold while capping conditional communication.

% \CorrectionModeFigure

\subsection{Request-Decoupled Pipeline}
\label{sec:cross-request}

\emph{Insight and approach.}
The next block of a request depends on the current block's committed prefix, whereas blocks from different requests have no prefix dependency on one another. Based on this observation, \method{} builds a request-decoupled pipeline that overlaps edge drafting, network transfer, and cloud verification across requests rather than speculating ahead within one request. It combines two mechanisms: confirmed-prefix scheduling admits at most one open block per request and switches the edge to another ready request while verification is pending; decoupled verification batching collects asynchronously arrived blocks in a global queue and dynamically re-batches them at the cloud, independently of their arrival grouping. Together, these mechanisms preserve per-request order and eliminate cross-block invalidation caused by unresolved prefixes while converting verification stalls and arrival skew into useful cross-request parallelism.

\subsubsection{Confirmed-Prefix Frontier}

A block is open from the start of drafting until it commits or aborts, and for
every request $\rho$, \methodsp enforces
\begin{equation}
N_{\mathrm{open}}(\rho)\leq 1.
\label{eq:openblock}
\end{equation}
A request is on the confirmed frontier only when it has no open block and both
KV cursors match its canonical prefix. It leaves the frontier when drafting
starts and returns only after the block resolves.

\emph{No cross-block invalidation.}
Under~\eqref{eq:openblock}, a rejection can invalidate only the uncommitted
suffix of the current block: no draft, transmission, or cloud verification for
a dependent successor exists. Waiting time is instead covered by another
frontier request whose prefix is already valid. This trades away
single-request overlap; when no other request is ready, the edge waits rather
than speculating from an unresolved prefix. Any fairness policy may choose
among ready requests without changing this property.

\subsubsection{Decoupled Verification Batching}

Requests produce blocks at different times because their draft depths, edge
service times, and network delays differ. Preserving their arrival grouping
would transfer this skew to the target verifier. \method{} instead places every
arrived block whose base prefix still matches the request's confirmed prefix in
a global verification queue and forms target batches from currently available
work.

This decoupling turns staggered cross-request arrivals into batching opportunities while keeping per-request order at commit. Conventional queue thresholds control launch timing without relaxing the confirmed-frontier or correction-fidelity invariants.

% \SchedulerBackpressureFigure

\FloatBarrier

\section{Evaluation}
\label{sec:evaluation}

\begin{figure*}[!t]
  \centering
  \includegraphics[width=0.994\textwidth]{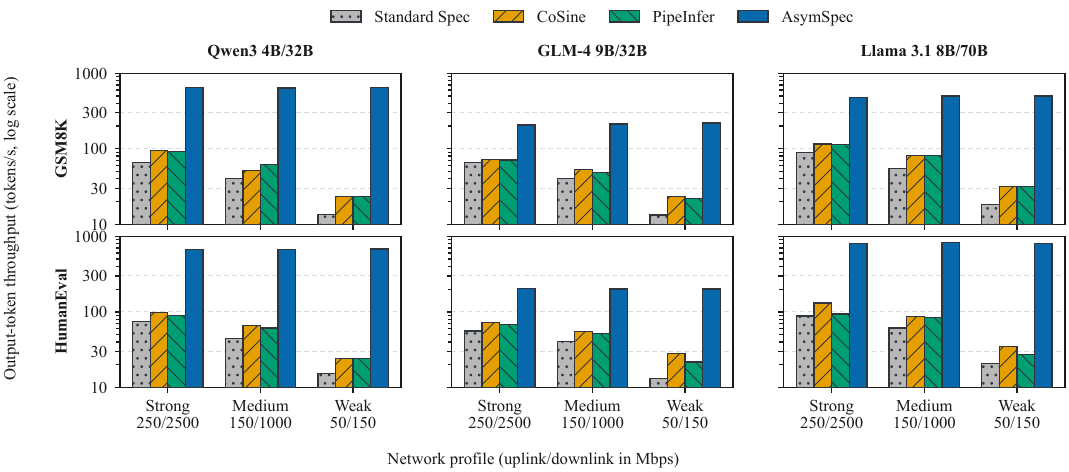}
  \caption{End-to-end output-token throughput under asymmetric network conditions. Each panel compares the four serving methods across the Strong, Medium, and Weak profiles; network labels report uplink/downlink bandwidth. Panels are organized by workload (rows) and draft/target model pair (columns). \method{} attains the highest throughput at all 18 operating points, up to 28.03$\times$ the throughput of the strongest baseline, and is the only method whose geometric-mean throughput does not drop from Strong to Weak. Bars share a logarithmic $y$-axis with a lower limit of 10 tokens/s; higher is better.}
  \label{fig:e2e-throughput}
\end{figure*}

\subsection{Evaluation Setup}

\emph{1) Implementation and Testbed:}
We implement \methodsp as two ZeroMQ-connected Python
services~\cite{hintjens2013zeromq}. The edge runs draft generation and request
scheduling on one RTX 3090; the cloud performs target verification and
independent re-batching on four A100 GPUs.

\emph{2) Models and Workloads:}
We evaluate three draft/target pairs: Qwen3 (4B/32B)~\cite{yang2025qwen3}, GLM-4
(9B/32B)~\cite{glm2024chatglm,zai2025glm40414}, and Llama 3.1
(8B/70B)~\cite{grattafiori2024llama}. These pairs span three model families
and 32B--70B targets. GSM8K~\cite{cobbe2021training} and
HumanEval~\cite{chen2021evaluating} represent mathematical reasoning and code
generation.

\emph{3) Network Conditions:}
Using Linux traffic control and NetEm~\cite{hemminger2005netem}, we emulate
three representative cloud--edge access scenarios with asymmetric
uplink/downlink capacity. Their bandwidth ranges are consistent with
measurements from commercial 4G/5G
networks~\cite{xu2020operational5g,kong2023accumo,narayanan2021variegated5g}.
{\setlength{\leftmargini}{1.25em}
\begin{itemize}
  \setlength{\itemsep}{0pt}
  \setlength{\parsep}{0pt}
  \setlength{\topsep}{2pt}
  \item \textbf{Weak---50/150 Mbps:} congested or LTE-like cellular access
  with scarce uplink capacity.
  \item \textbf{Medium---150/1,000 Mbps:} sub-6\,GHz 5G access with a faster
  downlink but still-limited uplink capacity.
  \item \textbf{Strong---250/2,500 Mbps:} high-capacity 5G/mmWave access,
  where network transfer is less likely to dominate.
\end{itemize}
}

\emph{4) Baselines:}
We compare \methodsp against three representative designs:
{\setlength{\leftmargini}{1.25em}
\begin{itemize}
  \setlength{\itemsep}{0pt}
  \setlength{\parsep}{0pt}
  \setlength{\topsep}{2pt}
  \item \textbf{Standard Spec}~\cite{leviathan2023,chen2023}: canonical
  speculative decoding and the algorithmic reference.
  \item \textbf{CoSine}~\cite{cosine}: collaborative cloud--edge speculative
  inference with decoupled execution.
  \item \textbf{PipeInfer}~\cite{pipeinfer}: asynchronous pipelined
  speculative inference.
\end{itemize}
}

\emph{5) Metrics:}
We report output-token throughput (tokens/s), end-to-end latency, time to first token (TTFT), and time per output token (TPOT).

\subsection{End-to-End Throughput}

\method{} attains the highest output-token throughput at all 18 model--workload--network operating points in Figure~\ref{fig:e2e-throughput}, achieving 2.82--28.03$\times$ the throughput of the strongest baseline at each point (geometric mean, 7.96$\times$). Geometrically averaged over the three network profiles, this ratio ranges from 4.20$\times$ for GLM-4 9B/32B on HumanEval to 12.55$\times$ for Qwen3 4B/32B on GSM8K. Across the 18 operating points, \method{} sustains 201.31--823.65 output tokens/s. Its geometric-mean throughput across the six model--workload pairs is 438.17, 443.81, and 446.36 tokens/s under the Strong, Medium, and Weak profiles, respectively.

We next examine how each method's throughput changes as the network degrades. From Strong to Weak, geometric-mean throughput falls by 78.6\% for Standard Spec, 71.4\% for CoSine, and 71.5\% for PipeInfer: every baseline loses roughly three-quarters of its throughput as the profile changes from 250/2,500 to 50/150\,Mbps. \method{} instead increases by 1.9\% over the same range. This pattern is consistent with its acceptance-sufficient common path and rejection-only downlink correction. Under the tested profiles, \method{}'s throughput is therefore much less sensitive to network degradation than that of the baselines. 

Differences across model--workload pairs reflect draft cost, acceptance behavior, and available batching work. Across the tested profiles, \method{} maintains stable geometric-mean throughput from Strong to Weak, unlike the large declines observed for all three baselines.

\subsection{Component Analysis}

\begin{table*}[!t]
\centering
\caption{Component analysis under three asymmetric network profiles. Entries
report absolute values followed by values relative to the complete \method{}
configuration, labeled Full in this table, in parentheses. Arrows indicate the
preferred direction; bold and underlining mark the best and second-best results
in each metric--network group.}
\label{tab:component-analysis}
\begingroup
\footnotesize
\setlength{\tabcolsep}{5pt}
\renewcommand{\arraystretch}{1.00}
\begin{tabular}{@{}ll@{\hspace{8pt}}ccc@{}}
\toprule
\multirow{2}{*}{\textbf{Metric}} &
\multirow{2}{*}{\textbf{Configuration}} &
\multicolumn{3}{c}{\textbf{Network profile (uplink/downlink)}} \\
\cmidrule(lr){3-5}
& & \shortstack{\textbf{Strong}\\250/2,500 Mbps} &
\shortstack{\textbf{Medium}\\150/1,000 Mbps} &
\shortstack{\textbf{Weak}\\50/150 Mbps} \\
\midrule
\multirow{5}{*}{\shortstack[l]{\textbf{Throughput}\\tokens/s $\uparrow$}} &
\textbf{Full} & \textbf{751.02 (100.0\%)} & \textbf{760.14 (100.0\%)} & \textbf{749.87 (100.0\%)} \\
\cmidrule(l){2-5}
& No asymmetric correction / top-$K$ & 47.46 (6.3\%) & 30.42 (4.0\%) & 10.15 (1.4\%) \\
& No top-$K$ & 559.45 (74.5\%) & 499.29 (65.7\%) & 179.59 (23.9\%) \\
& No pipeline & \underline{664.62} (88.5\%) & \underline{632.72} (83.2\%) & \underline{637.06} (85.0\%) \\
& No pipeline + runahead & 572.99 (76.3\%) & 580.88 (76.4\%) & 571.91 (76.3\%) \\
\midrule
\multirow{5}{*}{\shortstack[l]{\textbf{E2E latency}\\seconds $\downarrow$}} &
\textbf{Full} & \textbf{19.347 (1.00$\times$)} & \textbf{19.217 (1.00$\times$)} & \textbf{19.556 (1.00$\times$)} \\
\cmidrule(l){2-5}
& No asymmetric correction / top-$K$ & 621.154 (32.11$\times$) & 968.581 (50.40$\times$) & 2,966.391 (151.69$\times$) \\
& No top-$K$ & 38.281 (1.98$\times$) & 44.461 (2.31$\times$) & 149.660 (7.65$\times$) \\
& No pipeline & \underline{23.725} (1.23$\times$) & \underline{25.932} (1.35$\times$) & \underline{25.775} (1.32$\times$) \\
& No pipeline + runahead & 33.360 (1.72$\times$) & 32.677 (1.70$\times$) & 32.736 (1.67$\times$) \\
\midrule
\multirow{5}{*}{\shortstack[l]{\textbf{TTFT}\\seconds $\downarrow$}} &
\textbf{Full} & \textbf{0.432 (1.00$\times$)} & \textbf{0.434 (1.00$\times$)} & \textbf{0.443 (1.00$\times$)} \\
\cmidrule(l){2-5}
& No asymmetric correction / top-$K$ & 7.044 (16.32$\times$) & 11.826 (27.27$\times$) & 38.801 (87.64$\times$) \\
& No top-$K$ & 1.062 (2.46$\times$) & 1.151 (2.66$\times$) & 1.482 (3.35$\times$) \\
& No pipeline & \underline{0.531} (1.23$\times$) & \underline{0.627} (1.45$\times$) & 0.616 (1.39$\times$) \\
& No pipeline + runahead & 0.651 (1.51$\times$) & 0.640 (1.48$\times$) & \underline{0.565} (1.28$\times$) \\
\midrule
\multirow{5}{*}{\shortstack[l]{\textbf{TPOT}\\seconds $\downarrow$}} &
\textbf{Full} & \textbf{0.074 (1.00$\times$)} & \textbf{0.069 (1.00$\times$)} & \textbf{0.078 (1.00$\times$)} \\
\cmidrule(l){2-5}
& No asymmetric correction / top-$K$ & 2.408 (32.47$\times$) & 3.752 (54.30$\times$) & 11.481 (147.95$\times$) \\
& No top-$K$ & 0.146 (1.97$\times$) & 0.170 (2.46$\times$) & 0.581 (7.49$\times$) \\
& No pipeline & \underline{0.091} (1.23$\times$) & \underline{0.099} (1.44$\times$) & \underline{0.099} (1.27$\times$) \\
& No pipeline + runahead & 0.128 (1.73$\times$) & 0.126 (1.82$\times$) & 0.126 (1.63$\times$) \\
\bottomrule
\end{tabular}
\endgroup
\end{table*}

\begin{figure}[t]
  \centering
  \includegraphics[width=0.952\columnwidth]{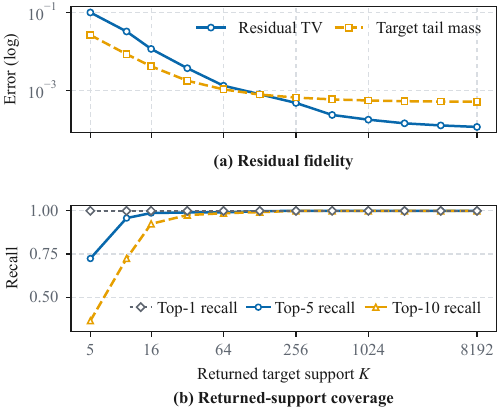}
  \caption{Correction fidelity on GSM8K with a Qwen3 4B draft and 32B target.
  Panel (a) reports residual TV and target-tail mass; (b) reports top-1, top-5,
  and top-10 recall within the returned support.}
  \label{fig:correction-fidelity}
\end{figure}

Table~\ref{tab:component-analysis} compares the Full configuration with variants that remove or replace asymmetric correction, top-$K$ support, and request-decoupled pipelining. Across the recorded aggregates, the Full configuration has the highest throughput and the lowest end-to-end latency, TTFT, and TPOT under all three network profiles. Even against the strongest reduced configuration in each profile, Full raises throughput by 13.0--20.1\% while also reducing end-to-end latency, TTFT, and TPOT.

The communication-side mechanisms become increasingly important as bandwidth declines. Jointly removing asymmetric correction and top-$K$ support reduces throughput by 93.7\%, 96.0\%, and 98.6\% relative to Full from Strong to Weak. The same configuration raises end-to-end latency to 32.1$\times$, 50.4$\times$, and 152$\times$; its TTFT and TPOT overheads similarly reach 87.6$\times$ and 148$\times$ in the Weak profile. Removing top-$K$ alone is less severe but still reduces throughput by 25.5--76.1\% and raises Weak-profile end-to-end latency to 7.7$\times$. Across these configurations, bounded correction support is associated with lower startup and token-generation overhead as bandwidth declines.

\begin{figure}[t]
  \centering
  \includegraphics[width=0.925\columnwidth]{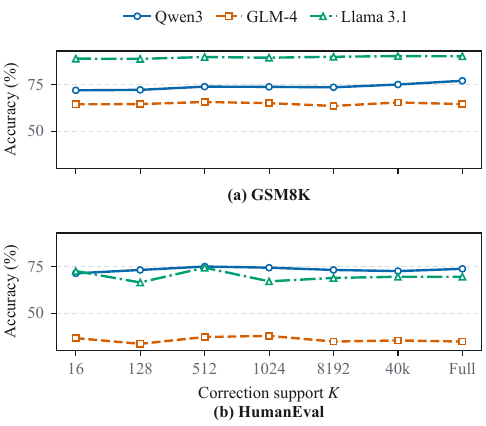}
  \caption{Task accuracy across correction-support settings on (a) GSM8K and
  (b) HumanEval. Lines identify the three draft/target model pairs.}
  \label{fig:correction-accuracy}
\end{figure}

\begin{figure*}[t]
  \centering
  \includegraphics[width=0.980\textwidth]{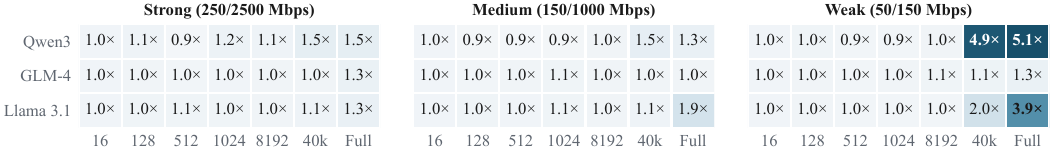}
  \caption{End-to-end latency for 100 requests at 4 requests/s. Each row is normalized independently to $K=16$ for the same model and network profile; labels give ratios rounded to one decimal place. Full denotes full-vocabulary correction.}
  \label{fig:correction-latency}
\end{figure*}

The scheduling variants expose a complementary effect. Removing the decoupled
pipeline lowers throughput by 11.5--16.8\% and raises the three latency metrics
to 1.2--1.4$\times$ Full. Replacing it with runahead retains only about 76\%
of Full's throughput and raises the same metrics to 1.3--1.8$\times$ Full, so
same-request runahead alone cannot recover the benefit of admitting independent work
across requests. Together, the throughput and latency columns in
Table~\ref{tab:component-analysis} show that asymmetric correction and top-$K$
control address communication amplification, whereas the decoupled pipeline
converts the resulting slack into useful cloud-side concurrency. The
ablation effects need not add linearly: changing correction traffic also changes
block arrival times and the amount of work available for a verification batch,
while changing the scheduler alters when link capacity is exercised. The
component study therefore supports a division of roles rather than an exact
numerical decomposition of total speedup. Communication control is most visible
as the network weakens; request decoupling remains useful across profiles
because arrival skew and per-request dependency persist even when bandwidth is
less restrictive.

\subsection{Correction Sensitivity}

The returned support $K$ trades correction payload against residual fidelity.
We examine its effect on protocol fidelity, task accuracy, and end-to-end
latency. The fidelity experiment uses a denser grid between $K=5$ and
$K=8{,}192$, while the accuracy and latency experiments use seven settings
from $K=16$ through Full.

\subsubsection{Protocol Fidelity}

Figure~\ref{fig:correction-fidelity} evaluates the Qwen3 4B draft and 32B target
pair on GSM8K. Residual TV and target-tail mass decrease rapidly before
$K=256$ and then flatten. Top-1 recall is 1.0 throughout the tested range from
$K=5$ to $K=8{,}192$. At $K=16$, top-5 and top-10 recall already reach 0.987
and 0.925, with a residual TV of 0.0116. At $K=256$, both recall values reach
1.0 and residual TV falls to $4.85\times10^{-4}$. Since the Qwen3 32B target
has 151,936 vocabulary entries, this support provides $593.5\times$
compression. Expanding to $K=8{,}192$ reduces residual TV by only another
$3.67\times10^{-4}$. Even at $K=8{,}192$, the returned support provides $18.55\times$ compression relative to the full 151,936-entry vocabulary. Thus, compact support already provides high correction
fidelity with a substantially smaller payload.

\subsubsection{Task Quality}
Figure~\ref{fig:correction-accuracy} compares bounded correction support with Full for each model and workload. Across the 36 within-pair comparisons, bounded support has an unweighted mean difference of $-0.52$ percentage points relative to Full. Separated by workload, the mean bounded-minus-Full difference is $-1.35$ percentage points on GSM8K and $+0.30$ percentage points on HumanEval. At least one bounded setting matches or exceeds Full in five of the six model--workload pairs; in the remaining Qwen3--GSM8K case, the best bounded setting achieves 75.0\% versus 77.0\% with Full. The direction of the pointwise difference varies across $K$, models, and workloads.

\subsubsection{Runtime Sensitivity}

Figure~\ref{fig:correction-latency} normalizes each row independently to the $K=16$ latency of the same model and network profile. The bounded settings from $K=128$ to $K=8{,}192$ remain close to this reference, spanning $0.88$--$1.19\times$ for Qwen3, $0.95$--$1.06\times$ for GLM-4, and $0.95$--$1.06\times$ for Llama 3.1. The $K=40{,}000$ setting is non-monotonic: for Qwen3 it reaches $1.50\times$, $1.47\times$, and $4.93\times$ under Strong, Medium, and Weak, respectively, while Llama 3.1 reaches $2.01\times$ under Weak. Full support becomes most expensive under the Weak profile, where latency rises to $5.1\times$, $1.3\times$, and $3.9\times$ the respective per-model references. Thus, supports up to $K=8{,}192$ add little latency within each model configuration, whereas $K=40{,}000$ and full-vocabulary support produce larger overheads in specific model--network combinations, most prominently for Qwen3 under Weak.

\section{Related Work}

\textbf{Speculative decoding.}
Classical speculative decoding uses a small draft model to propose tokens and a
larger target model to verify them in parallel while preserving the target
distribution~\cite{leviathan2023,chen2023}. Later work improves proposal
quality and verification efficiency through tree-structured candidates,
multi-token heads, and adaptive draft
structures~\cite{specinfer,medusa,sequoia}. Tree structures expose several
candidate continuations to one verification pass, multi-token heads predict
multiple future positions without repeatedly invoking a separate drafter, and
adaptive structures vary the candidate topology with the expected verification
gain. These methods primarily reduce target-model invocations within a local
inference stack; \method{} instead addresses the communication and dependency
costs that arise when drafting and verification run on opposite sides of an
asymmetric network.

\textbf{Cloud--edge collaborative inference.}
Recent systems place draft generation at the edge and target verification in
the cloud~\cite{dssd,specedge,pipesd,zhengcomm,zhengfast}. The verification
path runs through every speculative block, but cloud--edge bandwidth is
asymmetric: compact acceptance inputs repeatedly consume the constrained
uplink, whereas richer correction information is needed only after rejection
and can use the stronger downlink. Prior work sparsifies or adaptively
compresses probability messages~\cite{zhengcomm,sqssd}, while
Quantize-Sample-and-Verify offers an exact communication
mechanism~\cite{qsv}. \method{} is complementary:
it tests whether the returned target support is sufficient for the
\emph{realized residual}, composes bounded correction error across a request,
and falls back to exact recovery when the certificate fails.

\textbf{Serving and pipelining.}
Continuous batching and iteration-level scheduling improve accelerator
utilization for centralized LLM serving by admitting and retiring requests at
fine granularity, interleaving prefill and decode work, or assigning these
phases to different resources~\cite{vllm,sarathi,distserve}.
Distributed speculative systems further overlap drafting and verification,
interleave requests, or speculate beyond unresolved
blocks~\cite{pipeinfer,amusd,pearl,specedge,cosine,wisp}. In contrast, \method{}
permits at most one open block per request and fills verification waits only
with confirmed-prefix work from independent requests, eliminating
same-request invalidation cascades by construction.

\section{Conclusion}

\method{} combines certified progressive correction with a confirmed-prefix pipeline for cloud--edge speculative decoding. It keeps the common acceptance path compact, expands correction support only after rejection, and retains exact fallback; meanwhile, independent requests fill verification waits without same-request runahead. Bounded supports through $K=8{,}192$ closely track the task quality of full-vocabulary correction and the corresponding per-model, per-network $K=16$ latency reference, whereas very large or full-vocabulary supports can become costly on a weak downlink. Across three model pairs, two workloads, and three asymmetric network profiles, \method{} achieves \textbf{2.82--28.03$\times$} the output-token throughput of the strongest baseline, while its geometric-mean throughput changes by 1.9\% from Strong to Weak across the tested profiles.

\begingroup
% Keep each bibliography item intact across columns and pages.
\newcommand{\BIBdecl}{\interlinepenalty=10000\relax}
\bibliographystyle{IEEEtran}
\bibliography{asym_spec_refs}
\endgroup

\end{document}